\documentclass[apl,superscriptaddress,reprint,twocolumn,amssymb,aps,floatfix,showkeys]{revtex4-1}
\usepackage{amsmath}%
\usepackage{amsfonts}%
\usepackage{amssymb}%
\usepackage[pdftex]{graphicx}
\usepackage{color}
\usepackage{tabularx}
\usepackage{braket}
\usepackage{wasysym}
\usepackage[update,prepend]{epstopdf}

\newcommand{\ert}{Er$^{3+}$}

\begin{document}
\title{Evidence of Dilute Ferromagnetism in Rare-Earth doped YAG}

\author{Warrick G. Farr}
\affiliation{ARC Centre of Excellence for Engineered Quantum Systems, University of Western Australia, 35 Stirling Highway, Crawley WA 6009, Australia}

\author{Maxim Goryachev}
\affiliation{ARC Centre of Excellence for Engineered Quantum Systems, University of Western Australia, 35 Stirling Highway, Crawley WA 6009, Australia}

\author{Jean-Michel le Floch}
\affiliation{ARC Centre of Excellence for Engineered Quantum Systems, University of Western Australia, 35 Stirling Highway, Crawley WA 6009, Australia}

\author{Pavel Bushev}
\affiliation{Experimentalphysik, Universit\"{a}t des Saarlandes, D-66123 Saarbr\"{u}cken, Germany}

\author{Michael E. Tobar}
\affiliation{ARC Centre of Excellence for Engineered Quantum Systems, University of Western Australia, 35 Stirling Highway, Crawley WA 6009, Australia}


\begin{abstract}
This work demonstrates strong coupling regime between an Erbium ion spin ensemble and microwave Hybrid Cavity-Whispering Gallery Modes in a Yttrium Aluminium Garnet dielectric crystal. Coupling strengths of $220$~MHz and mode quality factors in excess of $10^6$ are demonstrated. Moreover, the magnetic response of high-Q modes demonstrates behaviour which is unusual for paramagnetic systems. This behaviour includes hysteresis and memory effects. {\color{black}Such qualitative change of the system's magnetic field response is interpreted as a phase transition of Rare Earth ion impurities. This phenomenon is similar to the phenomenon of dilute ferromagnetism in semiconductors. The clear temperature dependence of the phenomenon is demonstrated.}


\end{abstract}
\date{\today}
\maketitle






{\color{black}Quantum Electrodynamics (QED) investigates the quantum weirdness of the `light-matter' interactions and attempts to apply it to such emergence areas as quantum computing and quantum communications. For these purposes, research experiments utilize electromagnetic radiation (optical and microwave photons) and `matter' (spin ensembles, single ions, artificial superconducting qubits, etc).  One of the promising platforms for QED is known as `spins-in-solids' where a host for quantum matter is bounded within a dielectric crystal, which also serves as a 3D cavity. This particular type of a system may respectively serve as a quantum optical-to-microwave converter, an important element of the future quantum internet. Although, some feature of this apparatus have been already demonstrated\cite{PhysRevLett.105.140502,Schuster:2010rm,PhysRevLett.107.060502,PhysRevLett.110.157001,PhysRevB.90.075112,Abe:2011aa,Ranjan:2013aa}, the final goal is still out of reach of the present day technology and further investigation involving different types of crystals and atoms are required. In this letter, we demonstrate a `spins-in-solids' QED system based on Erbium ions in Yttrium Aluminium Garnet and an unusual phenomenon that has been explained by a dilute ferromagnetism phase transition.}


One such technology is based on Whispering Gallery Mode (WGM) single crystal resonators that serve simultaneously as photonic cavities and a spin system host. This unique feature delivers the highest possible magnetic filling factor with the lowest possible loss. In order to realise the full potential of this approach, it is important to find the best combination of the low loss crystal host and ion ensemble type and concentration (i.e. to optimise coherence time and photon-spin coupling). Examples of such systems vary from very dilute impurity ensembles of Iron Group Ions in Sapphire\cite{PhysRevB.88.224426, PhysRevA.89.013810} and Y$_2$SiO$_5$(YSO)\cite{Goryachev:2015aa}, Rare Earth Ions in YSO\cite{PhysRevLett.110.157001} and YAlO$_3$ (YAP)\cite{PhysRevB.90.075112} exhibiting Curie or van Vleck paramagnetism to highly concentrated ordered crystals such as YIG\cite{PhysRevApplied.2.054002,Tabuchi:2014aa,Zhang:2014aa,Weiler:2013aa} demonstrating ferromagnetic behavior. While such extremes of the concentration parameters in the frame work of the cavity QED have been already investigated, any intermediate regimes are still "terra incognita" despite the fact that spin-spin interactions lead to some extra observable effects {such as masing and other nonlinear effects}\cite{Bourhill2013b}.
 In this context it is important to study and understand implications of these spin-spin interactions and a corresponding magnetic phase for cavity QED in general and in particular quantum technology applications.

Here, we demonstrates an intermediate regime which can be qualified as `dilute ferromagnetism'\cite{RevModPhys.86.187,Dietl00,Dietl:2003aa}, where the long range order is established by dilute randomly distributed impurities in a non-magnetic crystal. {\color{black}Moreover, unlike all of the works on dilute ferromagnetism where the phenomenon is created by Iron Group ions, we demonstrate an experiment with an ensemble of Rare Earth ions, in particular erbium impurities in single crystal Yttrium Aluminium Garnet, Y$_3$Al$_5$O$_{12}$ (YAG). }
This combination of the impurity ensemble and the host with excellent optical properties\cite{LiuYAG} is interesting due to simultaneous existence of anisotropic microwave (X-band\cite{PF99} (8-12~GHz)) and optical (telecommunication band\cite{raey}) transition that can be utilised for quantum frequency converters\cite{PhysRevLett.113.063603,Williamson:2014aa,PhysRevLett.92.247902,PhysRevLett.105.220501,PhysRevA.87.052333,PhysRevA.80.033810}. 

\begin{figure}
	\centering
		\includegraphics[width=.99\linewidth]{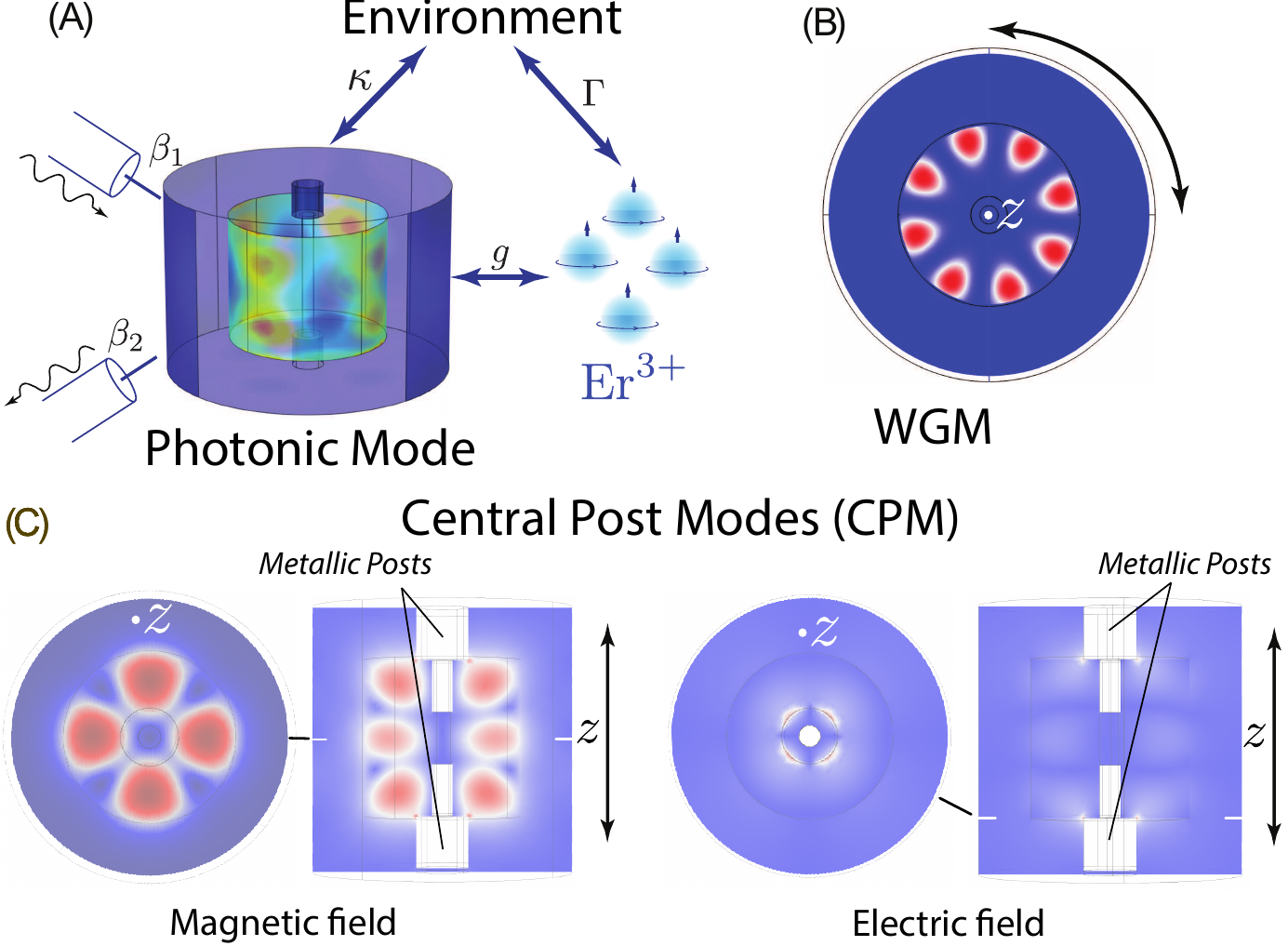}
	\caption{(A) Couplings between the subsystems: photonic cavity mode, \ert\ ensemble, the environment as well the external excitation and measurement apparatus. (B) A crystal WGM is created by the wave propagating along the cylinder circumference. (C) A Central Post Mode is characterised by strong field associated with the central post.}
		\label{fig:couplingplaceholder}
\end{figure}

The demonstration is performed with a QED system composed of a number of photonic modes and rare-earth spin ensemble (see Fig.~\ref{fig:couplingplaceholder}, (A)) coupled between each other with strength $g$ and coupled to the environment as $\kappa$ and $\Gamma$ respectively. The modes are probed via microwave transmission lines with $\beta_1$ and $\beta_2$ couplings. 
The role of the photonics mode is played by WGMs of a single crystal YAG playing a role of a cylindrical high $Q$-factor dielectric cavity (see Fig.~\ref{fig:couplingplaceholder} for a schematic view). 
The crystal is a host for Er$^{3+}$ magnetic impurities substituting Y$^{3+}$ ions during the growth process. The YAG cavity ($14.95$~mm in height, $15.00$~mm in diameter, with crystal $z$-axis aligned with the cylinder axis) is supplied by the manufacturer (Scientific Materials, Inc.) without a specified \ert\ concentration. 
The crystal is held by a pair of metallic posts and housed in a oxygen free copper cavity. For cavity transmission measurements, coupling of microwave signal lines to the system modes is realised with two straight antennae, parallel to the z-axis. Details of similar WGM spectroscopy is discussed previous works\cite{PhysRevB.88.224426, goryachev2013giant,PhysRevB.90.054409}.

Dielectric cylindrical resonators are typically designed to exhibit WGMs (see Fig.~\ref{fig:couplingplaceholder},~(B)). These modes are formed by standing or travelling waves formed by continuous total internal reflection from the cylindrical border between two media, i.e. dielectric and vacuum. Although, if the dielectric cylinder is enclosed into a metallic cavity with a central post, it may exhibit another type of mode associated with the post itself. These Central Post Modes (CPM) are closely related to the so-called re-entrant cavity modes\cite{Le-Floch:2013aa}. For the actual experimental structure, CPMs have been modelled numerically and classified\cite{suppm} with the good agreement between experiment and simulation. Moreover, corresponding filling factors correlate with the observed linewidths as the latter is determined by the former. 
These cavity modes are expected to have lower $Q$-factors because the metallic walls of the cavity act as an imperfect mirror, resulting in a considerable field close to metal surface or considerable current in the surface itself. In consequence, this structure allows cavity modes with complex field structures\cite{mikehybridmode2001}.

\begin{figure}
	\centering
		\includegraphics[width=1\linewidth]{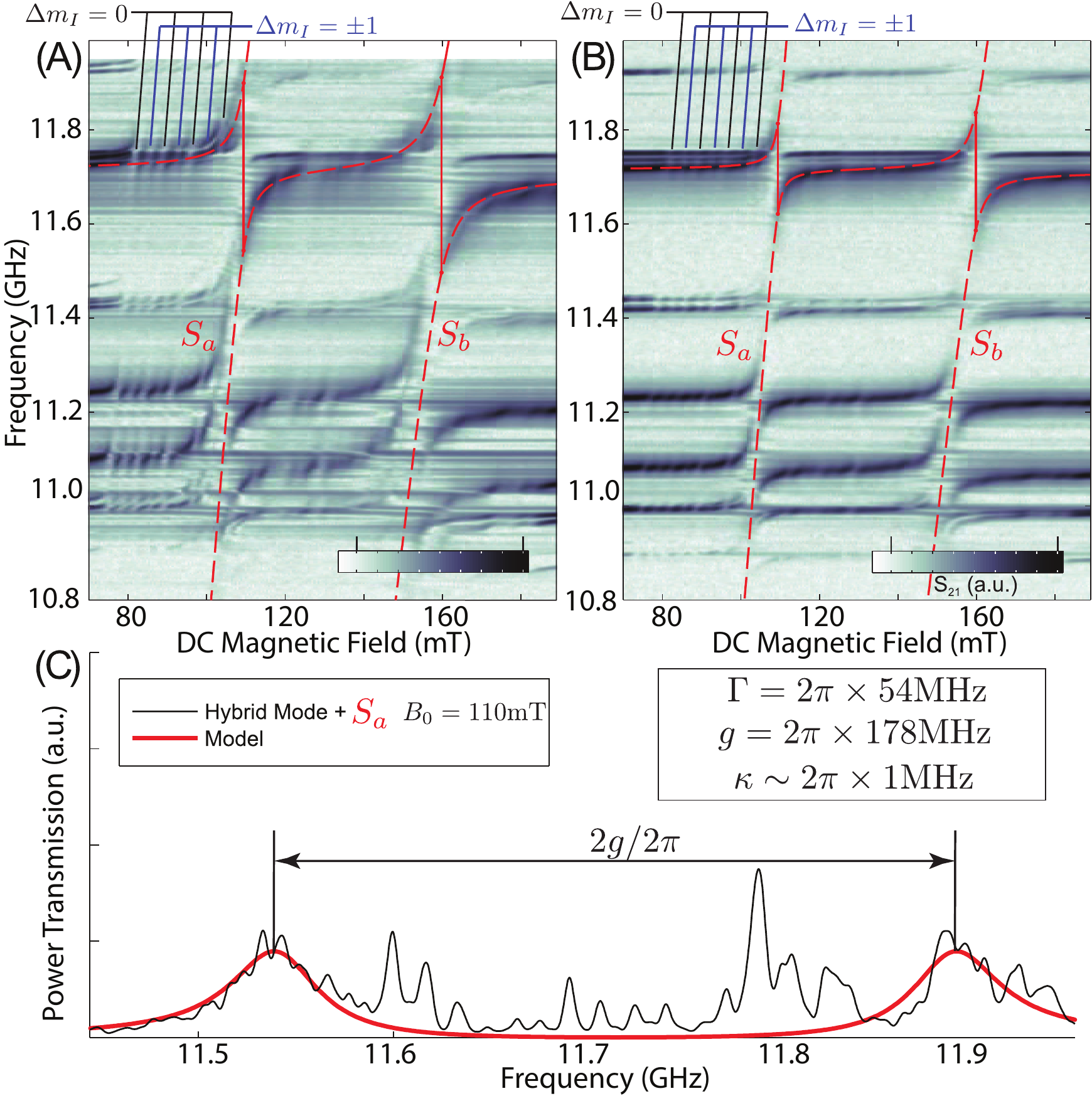}
		\caption{Transmission spectroscopy of the Er:YAG cavity showing a series of CPM modes exhibiting multiple interactions with the erbium ensemble for two temperatures: (A) $T=19$~mK, (B) $T=930$~mK. (C) Mode splitting for a $11.72$GHz CPM mode induced by the $S_a$ transition.}
	\label{fig:yagstrongcoup1}
\end{figure}

CPM modes have lower filling factors within the dielectric crystal, and thus exhibit relatively low values of $Q$-factor (in comparison to WGM) due to the significant dissipation in the central post. This results in mode linewidths $\kappa$ in the megahertz range which exceeds the linewidths of typical WGMs in YAG by several orders of magnitude\cite{0022-3727-35-13-301}.
In the low order CPM case, $\kappa>\Gamma$ for all the modes in the range 10.8 GHz and 12 GHz. The cavity transmission results depicting the CPM modes for two different values of system temperatures are shown in Fig.~\ref{fig:yagstrongcoup1}. Note that crystal WGMs with significantly higher $Q$-factors are not visible at this frequency scale. 

The darker regions in Fig.~\ref{fig:yagstrongcoup1} parallel to the magnetic field axis correspond to higher cavity transmission and represent CPM modes. When the DC magnetic field tunes the Er$^{3+}$ ensemble to a CPM mode frequency, the system exhibits an avoided level crossing (ALCs). There are clearly two major transitions of \ert, $S_a$ and $S_b$ with g-factors of $\text{g}_a=7.5$ and $\text{g}_b=5.16$. These lines correspond to the ESR $\Ket{-1/2}\rightarrow\Ket{+1/2}$ of \ert\ substituting for the Y$^{3+}$ at a single site {with D$_{2}$ symmetry, which has 2 subclasses of magnetically inequivalent sites \cite{raey,PhysRevB.77.085124}}. Coupling parameters of CPM modes in the $10-12$~GHz range are given in supplementary material\cite{suppm}. Despite comparatively large linewidths, several CPM modes exhibit strong coupling with both transitions of the erbium ensemble. This regime is characterised by the spin-photon coupling exceeding the cavity and spin ensemble linewidths $g>\kappa, \Gamma$. The strong coupling regime is primarily limited by the spin linewidth $\Gamma$ that is $2\pi\times54$ MHz.

The hyperfine structure of $^{167}$Er in both in Fig.~\ref{fig:yagstrongcoup1} (A) and (B) is most clearly visible at magnetic fields lower than the S$_{a}$ ensemble, since the lines otherwise overlap with the less well resolved hyperfine lines of the S$_{b}$ ensemble. Such hyperfine and quadrupole hyperfine structures have been previously observed in Er$^{3+}$ doped Yttrium Orthosilicate crystals at milli-Kelvin temperatures\cite{PhysRevLett.110.157001, PhysRevB.90.100404,PhysRevB.84.060501}. 

\begin{figure}
	\centering
		\includegraphics[width=.9\linewidth]{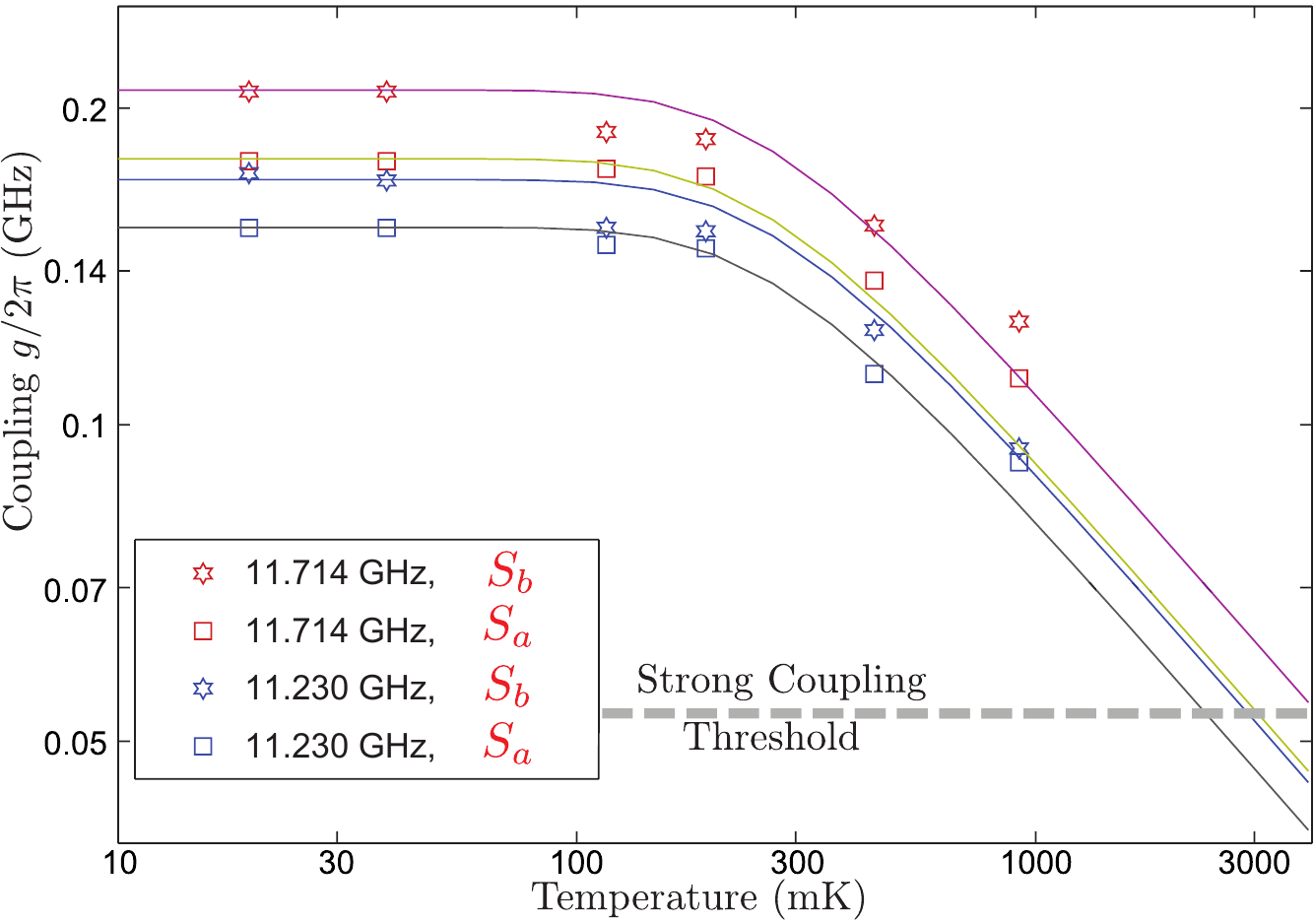}
	\caption{Temperature dependence of the coupling $g/2\pi$ between the selected CPM modes and electron spin transitions $S_a$, $S_b$. The dashed line shows $g/2\pi\sim54$ MHz  threshold for the strong coupling regime. {The solid curves show the theoretically predicted temperature dependence based on the model of independent spins.} }
	\label{fig:splitting_vs_Tgon2}
\end{figure}

Fig.~\ref{fig:splitting_vs_Tgon2} shows estimations of the coupling $g/2\pi$ of two selected CPM ($11.71$~GHz and $11.32$~GHz) to both Erbium principal transitions $S_a$ and $S_b$ as a function of environment temperature. {These data points are compared to the model of the independent spins\cite{Sandner:2012aa,PhysRevB.89.224407,PhysRevB.90.075112}}. The temperature dependence demonstrates a combination\cite{PhysRevB.89.224407} of the Curie linear dependence and van Vleck constant paramagnet dependencies above and below $200$mK respectively. The transition is due to the fact that below this temperature all electron spins have condensed to the lowest energy level $\ket{-1/2}$.
The strongest coupling occurs at the lowest temperatures between the $11.71$~GHz mode and the $S_b$ transition.

Coupling between a spin ensemble and a microwave cavity depend on a square root of the number of spins. For the experiment discussed in this work, {\it a priori} concentration of erbium impurities is not known. Although it can be deduced based on the knowledge of the measured coupling strength $g$ the photon filling factor $\xi_\perp$ using the following estimation: 
\begin{equation}
n=\frac{4 \hbar}{\omega_{0}\mu_0\xi_\perp}\left(\frac{g}{\text{g}_{AC}\mu_B}\right)^2,
\end{equation}
$\text{g}_{AC}$ is the AC g-factor, $\omega_0/2\pi$ is the resonance frequency and $\mu_B$ is the Bohr magnetron.
Whereas $g$ is determined from the experiment, $\xi_\perp$ is found with finite element modelling of the microwave cavity based on previously measured YAG permittivity\cite{krupka1999}. The resulting number {\color{black}$\xi_\perp\approx0.35$} describes the ratio of the magnetic field along the radial and axial directions within the crystal to the total magnetic field within the CPM mode. The density of the $S_a$ transition spins in the ground state is found from the interaction with the $\omega_0/2\pi=11.71$~GHz mode is {$\sim 9\times 10^{19}$~cm$^{-3}$.} The corresponding density of $S_b$ transition spins approaches {$\sim 2\times 10^{20}$ cm$^{-3}$.} These concentrations are significantly larger than typical concentrations used in other microwave QED experiments with rare-earth ion ensembles\cite{PhysRevLett.110.157001, PhysRevB.90.100404,PhysRevB.90.075112,Goryachev:2015aa}.  

For cavity QED experiments, it is imperative to have a low number of excitations. This parameter could be estimated based on the cavity incident power $P_{\text{inc}}$, cavity-transmission line couplings $\beta_1$ and $\beta_2$ and the system linewidth\cite{Hartnett:2011aa}:
 \begin{equation}
N=\frac{P_{\text{inc}}}{\hbar\omega_0\kappa}\frac{4\,\beta_1}{(1+\beta_1+\beta_2)^2}.
\end{equation}
giving mean number of photons in the cavity $N=13$ for the $11.71$~GHz mode and external couplings $\beta_1=\beta_2=10^{-4}$. {\color{black}This number can be characterised as a transition from the Tavis-Cummings model\cite{PhysRev.170.379,PhysRev.188.692}. It should be also mentioned that at the base temperature of 20mK, all cavity modes have thermal occupation number much less than one.}


Pure high-order WGMs exhibit a high confinement of the microwave energy inside the dielectric crystal, this results in significantly lower coupling to the environment, in particular to the metallic walls. Typical values of $Q$-factors for WGMs will exceed $10^5$ reaching $3\times 10^6$ for WGE$_{3,0,0}$, this inverts the relation between spin and photon couplings to the environment $\kappa\ll\Gamma$. The low linewidths of these modes allows us to resolve so-called mode doublets, degenerate modes whose reflection or time-reversal symmetries are broken due to various imperfections of the system\cite{PhysRevB.89.224407}. It should be mentioned that some  CPM modes are also capable of demonstrating large values of quality factor due to relatively good energy confinement. 
WGMs in $10-20$~GHz range were characterised in the same manner as the low $Q$-factor CPM modes, i.e. transmission measurements are made as a function of the external DC magnetic field. 

\begin{figure}
	\centering
		\includegraphics[width=.9\linewidth]{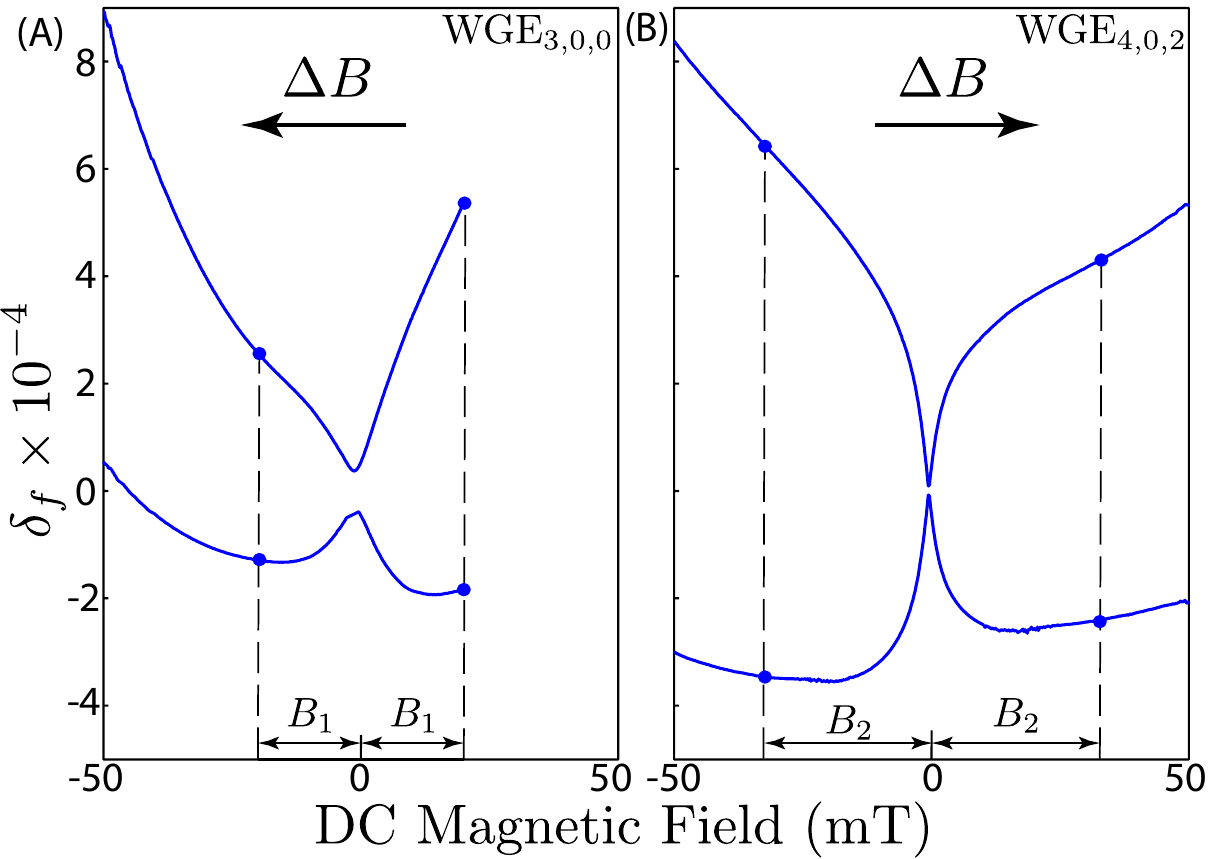}
\caption{Asymmetry of WGM doublets in terms of fractional frequency $\delta_f$ for two direction of the field change: (A) decreasing magnetic field ($\Delta B<0$), (B) increasing magnetic field ($\Delta B>0$). {The dashed lines show equal displacements of the external field in two opposite directions, which highlights the asymmetry.}}
\label{fig:Fra}
\end{figure}

{\color{black}The magnetic response of high-Q modes demonstrates {\it a memory effect}, which has not been observed in previous experiments with paramagnetic dielectrics at low temperatures.} This phenomenon reveals itself in the considerable difference between negative and positive DC magnetic fields scans as the magnetic field is progressively scanned and is dependent on the direction of change in magnetic field. {\color{black} The measurements are made over significant amount of time (order of days) in order to avoid heating effects due to changing magnetic field and to increase measurement sensitivity because of low excitation power levels.} {This asymmetry effect is demonstrated in Fig.~\ref{fig:Fra}, where the doublet fractional frequency shift $\delta_f=\frac{f-f_0}{f_0}$ of two independent modes are plotted against the applied magnetic field, where $f_0$ is the average resonance frequency of the doublet at zero field. The plots are obtained by scanning the field in opposite directions in two separate scans.} {\color{black} The possible explanation due to long spin T$_1$ times has been ruled out because very long measurement times were employed, very low excitation power that cannot saturate the ion ensemble and thus to substitute its behavior.} The memory phenomenon can also be explained with the hysteresis effect, which typically arises during the magnetisation of long-range-ordered magnetic materials such as ferromagnets. {It should be highlighted that this asymmetry has not been observed in any of the similar low doped crystals.}
Qualitative estimations of the observed hysteresis effects are given in supplementary material\cite{suppm}. It is demonstrated that the asymmetry for the higher frequency mode is always larger in magnitude. Additionally, the measurement output is dependent on magnetic field scan direction and prior history. {The hysteresis effect is reduced as the WGM resonance frequency is increased, because for the lower frequency modes are closer to the spin ensemble level splitting. Thus, there is a stronger impact of the Erbium ions on the cavity response near the zero external field and verifies the relationship between the hysteresis effect and the ion ensemble. }
Such hysteresis effects have not been observed with other paramagnetic crystals in similar experiments involving WGMs\cite{PhysRevB.88.224426, goryachev2013giant, PhysRevA.89.013810, PhysRevB.89.224407,Goryachev:2015aa} { or superconducting planar resonators\cite{Staudt:2012aa}, due to much lower concentration of the impurity ions}.  

\begin{figure}
	\centering
		\includegraphics[width=.9\linewidth]{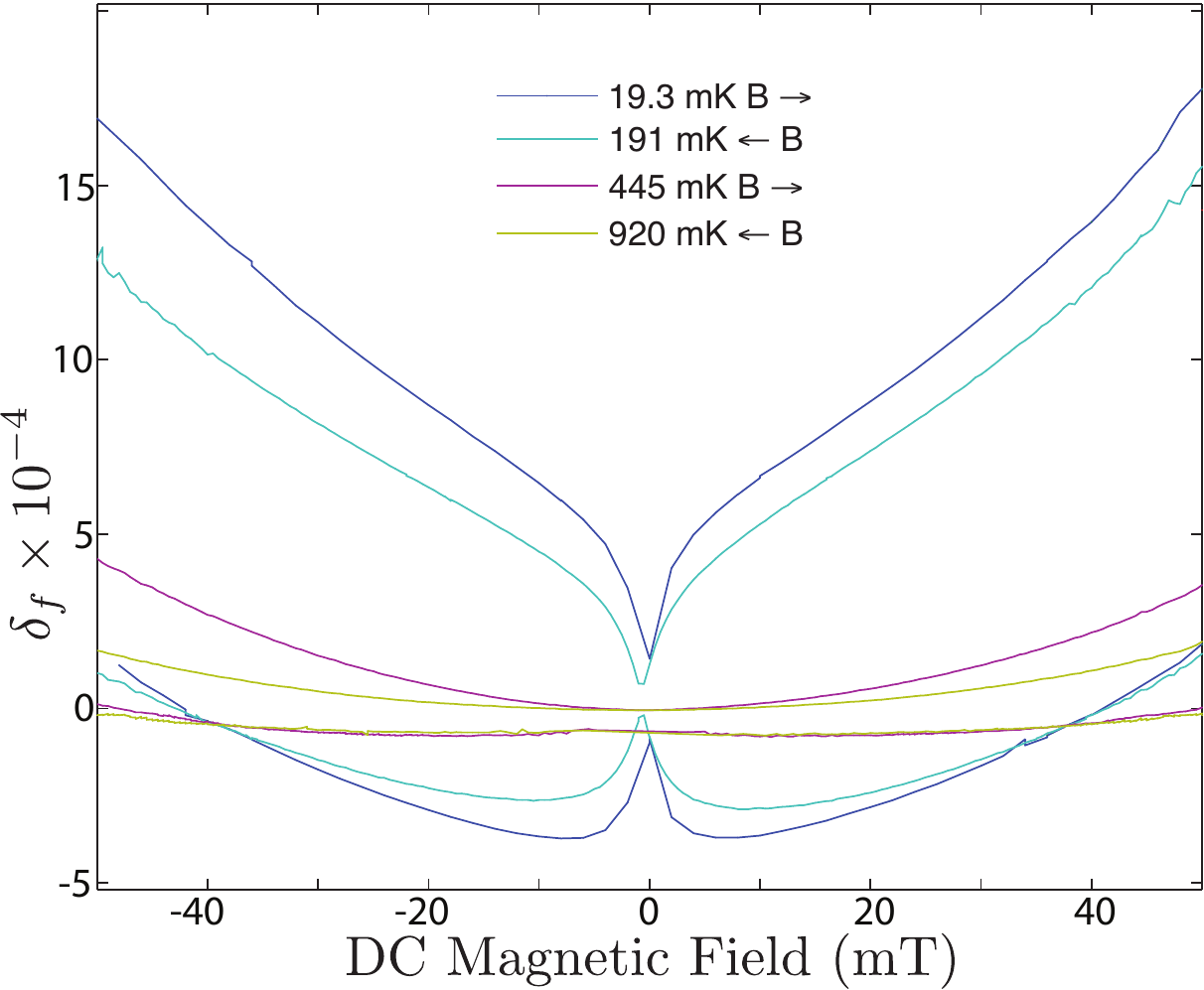}
\caption{Magnetic field response of WGM doublets in terms of the fractional frequency $\delta_f$. WGE$_{3,0,0}$ mode doublets measured at different temperatures. The qualitative change of behaviour happens between $190$~mK and $440$~mK.}
	\label{fig:YAGHIGHQ1editTEMP}
\end{figure} 

The WGE$_{3,0,0}$ mode was measured at a series of increasingly higher temperatures between $19$~mK and $930$~mK. The corresponding series of curves, shown in Fig.~\ref{fig:YAGHIGHQ1editTEMP} (B), indicates that  the system response qualitatively changes between $191$mK and $445$mK as confirmed by the $\eta_\pm$ factors listed in supplementary material\cite{suppm}. For example, the responses at $445$mK and $920$mK demonstrate negligible hysteresis and the bias field $B_0$ cannot be detected. This fact suggests that the impurity spin ensemble exhibits a phase transition between $191$ and $445$mK.

At lower temperatures the system demonstrates remanent magnetisation evident by the asymmetry of the WGM doublets. However, at higher temperatures, the spin-photon interaction behaves in the fully memoryless manner. { So, if the observed asymmetry, i.e. the memory effect and remanent magnetisation, is interpreted as the ferromagnetic phase, then the discussed transition in the system magnetic response represents the phase transition associated with a certain critical temperature between $191$ and $445$~mK. } This phase transition occurs when the average coupling between dilute impurity spins exceeds the thermal fluctuation energy leading to the long range order. Using the estimation for the spin ensemble density, one can find the average coupling between impurity ions\cite{Baibekov:2011aa} that, in the present case, approaches $J/\hbar\sim0.5$~GHz. Next, using the mean field approach, the critical temperature can be estimated as $T = \frac{zJ}{k_B}$ where $k_B$ is the Boltzmann constant and $z$ is the number of its nearest neighbours. So, in order to obtain the experimentally observed critical temperature above $0.2$K, $z$ should exceed $8.5$. Taking into account the relatively dense packing of erbium impurities of one per 8.5 unit cells, this number can be viewed as realistic. {\color{black} Note that some signature of ferromagnetic coupling between rare earth ions has been observed in weakly doped YVO$_4$, LiYF$_4$ and YAG where the effect of transition satellites is explained by dipole-dipole interaction between ion pairs\cite{PhysRevB.61.15338}}

In summary, we demonstrated a QED experiment with strong coupling between photons and erbium spin ensembles in YAG. The spin ensemble demonstrated behaviour unusual for paramagnetic systems, including remanent magnetisation, hysteresis and memory and sharp change of magnetic field response at a critical temperature in a dilute spin ensemble observed by virtue of WGMs. These facts draw analogies with the phenomenon of dilute ferromagnetism studied in semiconductors doped with Iron Group ions\cite{RevModPhys.86.187}. 


This work was supported by Australian Research Council grant CE110001013.


\begin{thebibliography}{45}%
\makeatletter
\providecommand \@ifxundefined [1]{%
 \@ifx{#1\undefined}
}%
\providecommand \@ifnum [1]{%
 \ifnum #1\expandafter \@firstoftwo
 \else \expandafter \@secondoftwo
 \fi
}%
\providecommand \@ifx [1]{%
 \ifx #1\expandafter \@firstoftwo
 \else \expandafter \@secondoftwo
 \fi
}%
\providecommand \natexlab [1]{#1}%
\providecommand \enquote  [1]{``#1''}%
\providecommand \bibnamefont  [1]{#1}%
\providecommand \bibfnamefont [1]{#1}%
\providecommand \citenamefont [1]{#1}%
\providecommand \href@noop [0]{\@secondoftwo}%
\providecommand \href [0]{\begingroup \@sanitize@url \@href}%
\providecommand \@href[1]{\@@startlink{#1}\@@href}%
\providecommand \@@href[1]{\endgroup#1\@@endlink}%
\providecommand \@sanitize@url [0]{\catcode `\\12\catcode `\$12\catcode
  `\&12\catcode `\#12\catcode `\^12\catcode `\_12\catcode `\%12\relax}%
\providecommand \@@startlink[1]{}%
\providecommand \@@endlink[0]{}%
\providecommand \url  [0]{\begingroup\@sanitize@url \@url }%
\providecommand \@url [1]{\endgroup\@href {#1}{\urlprefix }}%
\providecommand \urlprefix  [0]{URL }%
\providecommand \Eprint [0]{\href }%
\providecommand \doibase [0]{http://dx.doi.org/}%
\providecommand \selectlanguage [0]{\@gobble}%
\providecommand \bibinfo  [0]{\@secondoftwo}%
\providecommand \bibfield  [0]{\@secondoftwo}%
\providecommand \translation [1]{[#1]}%
\providecommand \BibitemOpen [0]{}%
\providecommand \bibitemStop [0]{}%
\providecommand \bibitemNoStop [0]{.\EOS\space}%
\providecommand \EOS [0]{\spacefactor3000\relax}%
\providecommand \BibitemShut  [1]{\csname bibitem#1\endcsname}%
\let\auto@bib@innerbib\@empty
\bibitem [{\citenamefont {Kubo}\ \emph {et~al.}(2010)\citenamefont {Kubo},
  \citenamefont {Ong}, \citenamefont {Bertet}, \citenamefont {Vion},
  \citenamefont {Jacques}, \citenamefont {Zheng}, \citenamefont {Dr\'eau},
  \citenamefont {Roch}, \citenamefont {Auffeves}, \citenamefont {Jelezko},
  \citenamefont {Wrachtrup}, \citenamefont {Barthe}, \citenamefont {Bergonzo},\
  and\ \citenamefont {Esteve}}]{PhysRevLett.105.140502}%
  \BibitemOpen
  \bibfield  {author} {\bibinfo {author} {\bibfnamefont {Y.}~\bibnamefont
  {Kubo}}, \bibinfo {author} {\bibfnamefont {F.~R.}\ \bibnamefont {Ong}},
  \bibinfo {author} {\bibfnamefont {P.}~\bibnamefont {Bertet}}, \bibinfo
  {author} {\bibfnamefont {D.}~\bibnamefont {Vion}}, \bibinfo {author}
  {\bibfnamefont {V.}~\bibnamefont {Jacques}}, \bibinfo {author} {\bibfnamefont
  {D.}~\bibnamefont {Zheng}}, \bibinfo {author} {\bibfnamefont
  {A.}~\bibnamefont {Dr\'eau}}, \bibinfo {author} {\bibfnamefont {J.-F.}\
  \bibnamefont {Roch}}, \bibinfo {author} {\bibfnamefont {A.}~\bibnamefont
  {Auffeves}}, \bibinfo {author} {\bibfnamefont {F.}~\bibnamefont {Jelezko}},
  \bibinfo {author} {\bibfnamefont {J.}~\bibnamefont {Wrachtrup}}, \bibinfo
  {author} {\bibfnamefont {M.~F.}\ \bibnamefont {Barthe}}, \bibinfo {author}
  {\bibfnamefont {P.}~\bibnamefont {Bergonzo}}, \ and\ \bibinfo {author}
  {\bibfnamefont {D.}~\bibnamefont {Esteve}},\ }\href {\doibase
  10.1103/PhysRevLett.105.140502} {\bibfield  {journal} {\bibinfo  {journal}
  {Phys. Rev. Lett.}\ }\textbf {\bibinfo {volume} {105}},\ \bibinfo {pages}
  {140502} (\bibinfo {year} {2010})}\BibitemShut {NoStop}%
\bibitem [{\citenamefont {Schuster}\ \emph {et~al.}(2010)\citenamefont
  {Schuster}, \citenamefont {Sears}, \citenamefont {Ginossar}, \citenamefont
  {DiCarlo}, \citenamefont {Frunzio}, \citenamefont {Morton}, \citenamefont
  {Wu}, \citenamefont {Briggs}, \citenamefont {Buckley}, \citenamefont
  {Awschalom},\ and\ \citenamefont {Schoelkopf}}]{Schuster:2010rm}%
  \BibitemOpen
  \bibfield  {author} {\bibinfo {author} {\bibfnamefont {D.~I.}\ \bibnamefont
  {Schuster}}, \bibinfo {author} {\bibfnamefont {A.~P.}\ \bibnamefont {Sears}},
  \bibinfo {author} {\bibfnamefont {E.}~\bibnamefont {Ginossar}}, \bibinfo
  {author} {\bibfnamefont {L.}~\bibnamefont {DiCarlo}}, \bibinfo {author}
  {\bibfnamefont {L.}~\bibnamefont {Frunzio}}, \bibinfo {author} {\bibfnamefont
  {J.~J.~L.}\ \bibnamefont {Morton}}, \bibinfo {author} {\bibfnamefont
  {H.}~\bibnamefont {Wu}}, \bibinfo {author} {\bibfnamefont {G.~A.~D.}\
  \bibnamefont {Briggs}}, \bibinfo {author} {\bibfnamefont {B.~B.}\
  \bibnamefont {Buckley}}, \bibinfo {author} {\bibfnamefont {D.~D.}\
  \bibnamefont {Awschalom}}, \ and\ \bibinfo {author} {\bibfnamefont {R.~J.}\
  \bibnamefont {Schoelkopf}},\ }\href
  {http://link.aps.org/doi/10.1103/PhysRevLett.105.140501} {\bibfield
  {journal} {\bibinfo  {journal} {Physical Review Letters}\ }\textbf {\bibinfo
  {volume} {105}},\ \bibinfo {pages} {140501} (\bibinfo {year}
  {2010})}\BibitemShut {NoStop}%
\bibitem [{\citenamefont {Ams\"uss}\ \emph {et~al.}(2011)\citenamefont
  {Ams\"uss}, \citenamefont {Koller}, \citenamefont {N\"obauer}, \citenamefont
  {Putz}, \citenamefont {Rotter}, \citenamefont {Sandner}, \citenamefont
  {Schneider}, \citenamefont {Schramb\"ock}, \citenamefont {Steinhauser},
  \citenamefont {Ritsch}, \citenamefont {Schmiedmayer},\ and\ \citenamefont
  {Majer}}]{PhysRevLett.107.060502}%
  \BibitemOpen
  \bibfield  {author} {\bibinfo {author} {\bibfnamefont {R.}~\bibnamefont
  {Ams\"uss}}, \bibinfo {author} {\bibfnamefont {C.}~\bibnamefont {Koller}},
  \bibinfo {author} {\bibfnamefont {T.}~\bibnamefont {N\"obauer}}, \bibinfo
  {author} {\bibfnamefont {S.}~\bibnamefont {Putz}}, \bibinfo {author}
  {\bibfnamefont {S.}~\bibnamefont {Rotter}}, \bibinfo {author} {\bibfnamefont
  {K.}~\bibnamefont {Sandner}}, \bibinfo {author} {\bibfnamefont
  {S.}~\bibnamefont {Schneider}}, \bibinfo {author} {\bibfnamefont
  {M.}~\bibnamefont {Schramb\"ock}}, \bibinfo {author} {\bibfnamefont
  {G.}~\bibnamefont {Steinhauser}}, \bibinfo {author} {\bibfnamefont
  {H.}~\bibnamefont {Ritsch}}, \bibinfo {author} {\bibfnamefont
  {J.}~\bibnamefont {Schmiedmayer}}, \ and\ \bibinfo {author} {\bibfnamefont
  {J.}~\bibnamefont {Majer}},\ }\href {\doibase 10.1103/PhysRevLett.107.060502}
  {\bibfield  {journal} {\bibinfo  {journal} {Phys. Rev. Lett.}\ }\textbf
  {\bibinfo {volume} {107}},\ \bibinfo {pages} {060502} (\bibinfo {year}
  {2011})}\BibitemShut {NoStop}%
\bibitem [{\citenamefont {Probst}\ \emph {et~al.}(2013)\citenamefont {Probst},
  \citenamefont {Rotzinger}, \citenamefont {W\"unsch}, \citenamefont {Jung},
  \citenamefont {Jerger}, \citenamefont {Siegel}, \citenamefont {Ustinov},\
  and\ \citenamefont {Bushev}}]{PhysRevLett.110.157001}%
  \BibitemOpen
  \bibfield  {author} {\bibinfo {author} {\bibfnamefont {S.}~\bibnamefont
  {Probst}}, \bibinfo {author} {\bibfnamefont {H.}~\bibnamefont {Rotzinger}},
  \bibinfo {author} {\bibfnamefont {S.}~\bibnamefont {W\"unsch}}, \bibinfo
  {author} {\bibfnamefont {P.}~\bibnamefont {Jung}}, \bibinfo {author}
  {\bibfnamefont {M.}~\bibnamefont {Jerger}}, \bibinfo {author} {\bibfnamefont
  {M.}~\bibnamefont {Siegel}}, \bibinfo {author} {\bibfnamefont {A.~V.}\
  \bibnamefont {Ustinov}}, \ and\ \bibinfo {author} {\bibfnamefont {P.~A.}\
  \bibnamefont {Bushev}},\ }\href {\doibase 10.1103/PhysRevLett.110.157001}
  {\bibfield  {journal} {\bibinfo  {journal} {Phys. Rev. Lett.}\ }\textbf
  {\bibinfo {volume} {110}},\ \bibinfo {pages} {157001} (\bibinfo {year}
  {2013})}\BibitemShut {NoStop}%
\bibitem [{\citenamefont {Tkal\ifmmode~\check{c}\else \v{c}\fi{}ec}\ \emph
  {et~al.}(2014)\citenamefont {Tkal\ifmmode~\check{c}\else \v{c}\fi{}ec},
  \citenamefont {Probst}, \citenamefont {Rieger}, \citenamefont {Rotzinger},
  \citenamefont {W\"unsch}, \citenamefont {Kukharchyk}, \citenamefont {Wieck},
  \citenamefont {Siegel}, \citenamefont {Ustinov},\ and\ \citenamefont
  {Bushev}}]{PhysRevB.90.075112}%
  \BibitemOpen
  \bibfield  {author} {\bibinfo {author} {\bibfnamefont {A.}~\bibnamefont
  {Tkal\ifmmode~\check{c}\else \v{c}\fi{}ec}}, \bibinfo {author} {\bibfnamefont
  {S.}~\bibnamefont {Probst}}, \bibinfo {author} {\bibfnamefont
  {D.}~\bibnamefont {Rieger}}, \bibinfo {author} {\bibfnamefont
  {H.}~\bibnamefont {Rotzinger}}, \bibinfo {author} {\bibfnamefont
  {S.}~\bibnamefont {W\"unsch}}, \bibinfo {author} {\bibfnamefont
  {N.}~\bibnamefont {Kukharchyk}}, \bibinfo {author} {\bibfnamefont {A.~D.}\
  \bibnamefont {Wieck}}, \bibinfo {author} {\bibfnamefont {M.}~\bibnamefont
  {Siegel}}, \bibinfo {author} {\bibfnamefont {A.~V.}\ \bibnamefont {Ustinov}},
  \ and\ \bibinfo {author} {\bibfnamefont {P.}~\bibnamefont {Bushev}},\ }\href
  {\doibase 10.1103/PhysRevB.90.075112} {\bibfield  {journal} {\bibinfo
  {journal} {Phys. Rev. B}\ }\textbf {\bibinfo {volume} {90}},\ \bibinfo
  {pages} {075112} (\bibinfo {year} {2014})}\BibitemShut {NoStop}%
\bibitem [{\citenamefont {Abe}\ \emph {et~al.}(2011)\citenamefont {Abe},
  \citenamefont {Wu}, \citenamefont {Ardavan},\ and\ \citenamefont
  {Morton}}]{Abe:2011aa}%
  \BibitemOpen
  \bibfield  {author} {\bibinfo {author} {\bibfnamefont {E.}~\bibnamefont
  {Abe}}, \bibinfo {author} {\bibfnamefont {H.}~\bibnamefont {Wu}}, \bibinfo
  {author} {\bibfnamefont {A.}~\bibnamefont {Ardavan}}, \ and\ \bibinfo
  {author} {\bibfnamefont {J.~J.~L.}\ \bibnamefont {Morton}},\ }\href
  {http://scitation.aip.org/content/aip/journal/apl/98/25/10.1063/1.3601930}
  {\bibfield  {journal} {\bibinfo  {journal} {Applied Physics Letters}\
  }\textbf {\bibinfo {volume} {98}},\ \bibinfo {pages} {251108} (\bibinfo
  {year} {2011})}\BibitemShut {NoStop}%
\bibitem [{\citenamefont {Ranjan}\ \emph {et~al.}(2013)\citenamefont {Ranjan},
  \citenamefont {de~Lange}, \citenamefont {Schutjens}, \citenamefont
  {Debelhoir}, \citenamefont {Groen}, \citenamefont {Szombati}, \citenamefont
  {Thoen}, \citenamefont {Klapwijk}, \citenamefont {Hanson},\ and\
  \citenamefont {DiCarlo}}]{Ranjan:2013aa}%
  \BibitemOpen
  \bibfield  {author} {\bibinfo {author} {\bibfnamefont {V.}~\bibnamefont
  {Ranjan}}, \bibinfo {author} {\bibfnamefont {G.}~\bibnamefont {de~Lange}},
  \bibinfo {author} {\bibfnamefont {R.}~\bibnamefont {Schutjens}}, \bibinfo
  {author} {\bibfnamefont {T.}~\bibnamefont {Debelhoir}}, \bibinfo {author}
  {\bibfnamefont {J.~P.}\ \bibnamefont {Groen}}, \bibinfo {author}
  {\bibfnamefont {D.}~\bibnamefont {Szombati}}, \bibinfo {author}
  {\bibfnamefont {D.~J.}\ \bibnamefont {Thoen}}, \bibinfo {author}
  {\bibfnamefont {T.~M.}\ \bibnamefont {Klapwijk}}, \bibinfo {author}
  {\bibfnamefont {R.}~\bibnamefont {Hanson}}, \ and\ \bibinfo {author}
  {\bibfnamefont {L.}~\bibnamefont {DiCarlo}},\ }\href
  {http://link.aps.org/doi/10.1103/PhysRevLett.110.067004} {\bibfield
  {journal} {\bibinfo  {journal} {Physical Review Letters}\ }\textbf {\bibinfo
  {volume} {110}},\ \bibinfo {pages} {067004} (\bibinfo {year}
  {2013})}\BibitemShut {NoStop}%
\bibitem [{\citenamefont {Farr}\ \emph {et~al.}(2013)\citenamefont {Farr},
  \citenamefont {Creedon}, \citenamefont {Goryachev}, \citenamefont
  {Benmessa\"\i},\ and\ \citenamefont {Tobar}}]{PhysRevB.88.224426}%
  \BibitemOpen
  \bibfield  {author} {\bibinfo {author} {\bibfnamefont {W.~G.}\ \bibnamefont
  {Farr}}, \bibinfo {author} {\bibfnamefont {D.~L.}\ \bibnamefont {Creedon}},
  \bibinfo {author} {\bibfnamefont {M.}~\bibnamefont {Goryachev}}, \bibinfo
  {author} {\bibfnamefont {K.}~\bibnamefont {Benmessa\"\i}}, \ and\ \bibinfo
  {author} {\bibfnamefont {M.~E.}\ \bibnamefont {Tobar}},\ }\href {\doibase
  10.1103/PhysRevB.88.224426} {\bibfield  {journal} {\bibinfo  {journal} {Phys.
  Rev. B}\ }\textbf {\bibinfo {volume} {88}},\ \bibinfo {pages} {224426}
  (\bibinfo {year} {2013})}\BibitemShut {NoStop}%
\bibitem [{\citenamefont {Goryachev}\ \emph
  {et~al.}(2014{\natexlab{a}})\citenamefont {Goryachev}, \citenamefont {Farr},
  \citenamefont {Creedon},\ and\ \citenamefont {Tobar}}]{PhysRevA.89.013810}%
  \BibitemOpen
  \bibfield  {author} {\bibinfo {author} {\bibfnamefont {M.}~\bibnamefont
  {Goryachev}}, \bibinfo {author} {\bibfnamefont {W.~G.}\ \bibnamefont {Farr}},
  \bibinfo {author} {\bibfnamefont {D.~L.}\ \bibnamefont {Creedon}}, \ and\
  \bibinfo {author} {\bibfnamefont {M.~E.}\ \bibnamefont {Tobar}},\ }\href
  {\doibase 10.1103/PhysRevA.89.013810} {\bibfield  {journal} {\bibinfo
  {journal} {Phys. Rev. A}\ }\textbf {\bibinfo {volume} {89}},\ \bibinfo
  {pages} {013810} (\bibinfo {year} {2014}{\natexlab{a}})}\BibitemShut
  {NoStop}%
\bibitem [{\citenamefont {Goryachev}\ \emph {et~al.}(2015)\citenamefont
  {Goryachev}, \citenamefont {Farr}, \citenamefont {Carmo~Carvalho},
  \citenamefont {Creedon}, \citenamefont {Le~Floch}, \citenamefont {Probst},
  \citenamefont {Bushev},\ and\ \citenamefont {Tobar}}]{Goryachev:2015aa}%
  \BibitemOpen
  \bibfield  {author} {\bibinfo {author} {\bibfnamefont {M.}~\bibnamefont
  {Goryachev}}, \bibinfo {author} {\bibfnamefont {W.~G.}\ \bibnamefont {Farr}},
  \bibinfo {author} {\bibfnamefont {N.~d.}\ \bibnamefont {Carmo~Carvalho}},
  \bibinfo {author} {\bibfnamefont {D.~L.}\ \bibnamefont {Creedon}}, \bibinfo
  {author} {\bibfnamefont {J.-M.}\ \bibnamefont {Le~Floch}}, \bibinfo {author}
  {\bibfnamefont {S.}~\bibnamefont {Probst}}, \bibinfo {author} {\bibfnamefont
  {P.}~\bibnamefont {Bushev}}, \ and\ \bibinfo {author} {\bibfnamefont {M.~E.}\
  \bibnamefont {Tobar}},\ }\href
  {http://scitation.aip.org/content/aip/journal/apl/106/23/10.1063/1.4922376}
  {\bibfield  {journal} {\bibinfo  {journal} {Applied Physics Letters}\
  }\textbf {\bibinfo {volume} {106}},\ \bibinfo {pages} {232401} (\bibinfo
  {year} {2015})}\BibitemShut {NoStop}%
\bibitem [{\citenamefont {Goryachev}\ \emph
  {et~al.}(2014{\natexlab{b}})\citenamefont {Goryachev}, \citenamefont {Farr},
  \citenamefont {Creedon}, \citenamefont {Fan}, \citenamefont {Kostylev},\ and\
  \citenamefont {Tobar}}]{PhysRevApplied.2.054002}%
  \BibitemOpen
  \bibfield  {author} {\bibinfo {author} {\bibfnamefont {M.}~\bibnamefont
  {Goryachev}}, \bibinfo {author} {\bibfnamefont {W.~G.}\ \bibnamefont {Farr}},
  \bibinfo {author} {\bibfnamefont {D.~L.}\ \bibnamefont {Creedon}}, \bibinfo
  {author} {\bibfnamefont {Y.}~\bibnamefont {Fan}}, \bibinfo {author}
  {\bibfnamefont {M.}~\bibnamefont {Kostylev}}, \ and\ \bibinfo {author}
  {\bibfnamefont {M.~E.}\ \bibnamefont {Tobar}},\ }\href {\doibase
  10.1103/PhysRevApplied.2.054002} {\bibfield  {journal} {\bibinfo  {journal}
  {Phys. Rev. Applied}\ }\textbf {\bibinfo {volume} {2}},\ \bibinfo {pages}
  {054002} (\bibinfo {year} {2014}{\natexlab{b}})}\BibitemShut {NoStop}%
\bibitem [{\citenamefont {Tabuchi}\ \emph {et~al.}(2014)\citenamefont
  {Tabuchi}, \citenamefont {Ishino}, \citenamefont {Ishikawa}, \citenamefont
  {Yamazaki}, \citenamefont {Usami},\ and\ \citenamefont
  {Nakamura}}]{Tabuchi:2014aa}%
  \BibitemOpen
  \bibfield  {author} {\bibinfo {author} {\bibfnamefont {Y.}~\bibnamefont
  {Tabuchi}}, \bibinfo {author} {\bibfnamefont {S.}~\bibnamefont {Ishino}},
  \bibinfo {author} {\bibfnamefont {T.}~\bibnamefont {Ishikawa}}, \bibinfo
  {author} {\bibfnamefont {R.}~\bibnamefont {Yamazaki}}, \bibinfo {author}
  {\bibfnamefont {K.}~\bibnamefont {Usami}}, \ and\ \bibinfo {author}
  {\bibfnamefont {Y.}~\bibnamefont {Nakamura}},\ }\href
  {http://link.aps.org/doi/10.1103/PhysRevLett.113.083603} {\bibfield
  {journal} {\bibinfo  {journal} {Physical Review Letters}\ }\textbf {\bibinfo
  {volume} {113}},\ \bibinfo {pages} {083603} (\bibinfo {year}
  {2014})}\BibitemShut {NoStop}%
\bibitem [{\citenamefont {Zhang}\ \emph {et~al.}(2014)\citenamefont {Zhang},
  \citenamefont {Zou}, \citenamefont {Jiang},\ and\ \citenamefont
  {Tang}}]{Zhang:2014aa}%
  \BibitemOpen
  \bibfield  {author} {\bibinfo {author} {\bibfnamefont {X.}~\bibnamefont
  {Zhang}}, \bibinfo {author} {\bibfnamefont {C.-L.}\ \bibnamefont {Zou}},
  \bibinfo {author} {\bibfnamefont {L.}~\bibnamefont {Jiang}}, \ and\ \bibinfo
  {author} {\bibfnamefont {H.~X.}\ \bibnamefont {Tang}},\ }\href
  {http://link.aps.org/doi/10.1103/PhysRevLett.113.156401} {\bibfield
  {journal} {\bibinfo  {journal} {Physical Review Letters}\ }\textbf {\bibinfo
  {volume} {113}},\ \bibinfo {pages} {156401} (\bibinfo {year}
  {2014})}\BibitemShut {NoStop}%
\bibitem [{\citenamefont {Weiler}\ \emph {et~al.}(2013)\citenamefont {Weiler},
  \citenamefont {Althammer}, \citenamefont {Schreier}, \citenamefont {Lotze},
  \citenamefont {Pernpeintner}, \citenamefont {Meyer}, \citenamefont {Huebl},
  \citenamefont {Gross}, \citenamefont {Kamra}, \citenamefont {Xiao},
  \citenamefont {Chen}, \citenamefont {Jiao}, \citenamefont {Bauer},\ and\
  \citenamefont {Goennenwein}}]{Weiler:2013aa}%
  \BibitemOpen
  \bibfield  {author} {\bibinfo {author} {\bibfnamefont {M.}~\bibnamefont
  {Weiler}}, \bibinfo {author} {\bibfnamefont {M.}~\bibnamefont {Althammer}},
  \bibinfo {author} {\bibfnamefont {M.}~\bibnamefont {Schreier}}, \bibinfo
  {author} {\bibfnamefont {J.}~\bibnamefont {Lotze}}, \bibinfo {author}
  {\bibfnamefont {M.}~\bibnamefont {Pernpeintner}}, \bibinfo {author}
  {\bibfnamefont {S.}~\bibnamefont {Meyer}}, \bibinfo {author} {\bibfnamefont
  {H.}~\bibnamefont {Huebl}}, \bibinfo {author} {\bibfnamefont
  {R.}~\bibnamefont {Gross}}, \bibinfo {author} {\bibfnamefont
  {A.}~\bibnamefont {Kamra}}, \bibinfo {author} {\bibfnamefont
  {J.}~\bibnamefont {Xiao}}, \bibinfo {author} {\bibfnamefont {Y.-T.}\
  \bibnamefont {Chen}}, \bibinfo {author} {\bibfnamefont {H.}~\bibnamefont
  {Jiao}}, \bibinfo {author} {\bibfnamefont {G.~E.~W.}\ \bibnamefont {Bauer}},
  \ and\ \bibinfo {author} {\bibfnamefont {S.~T.~B.}\ \bibnamefont
  {Goennenwein}},\ }\href
  {http://link.aps.org/doi/10.1103/PhysRevLett.111.176601} {\bibfield
  {journal} {\bibinfo  {journal} {Physical Review Letters}\ }\textbf {\bibinfo
  {volume} {111}},\ \bibinfo {pages} {176601} (\bibinfo {year}
  {2013})}\BibitemShut {NoStop}%
\bibitem [{\citenamefont {Bourhill}\ \emph {et~al.}(2013)\citenamefont
  {Bourhill}, \citenamefont {Benmessai}, \citenamefont {Goryachev},
  \citenamefont {Creedon}, \citenamefont {Farr},\ and\ \citenamefont
  {Tobar}}]{Bourhill2013b}%
  \BibitemOpen
  \bibfield  {author} {\bibinfo {author} {\bibfnamefont {J.}~\bibnamefont
  {Bourhill}}, \bibinfo {author} {\bibfnamefont {K.}~\bibnamefont {Benmessai}},
  \bibinfo {author} {\bibfnamefont {M.}~\bibnamefont {Goryachev}}, \bibinfo
  {author} {\bibfnamefont {D.~L.}\ \bibnamefont {Creedon}}, \bibinfo {author}
  {\bibfnamefont {W.}~\bibnamefont {Farr}}, \ and\ \bibinfo {author}
  {\bibfnamefont {M.~E.}\ \bibnamefont {Tobar}},\ }\href {\doibase
  10.1103/PhysRevB.88.235104} {\bibfield  {journal} {\bibinfo  {journal} {Phys.
  Rev. B}\ }\textbf {\bibinfo {volume} {88}},\ \bibinfo {pages} {235104}
  (\bibinfo {year} {2013})}\BibitemShut {NoStop}%
\bibitem [{\citenamefont {Dietl}\ and\ \citenamefont
  {Ohno}(2014)}]{RevModPhys.86.187}%
  \BibitemOpen
  \bibfield  {author} {\bibinfo {author} {\bibfnamefont {T.}~\bibnamefont
  {Dietl}}\ and\ \bibinfo {author} {\bibfnamefont {H.}~\bibnamefont {Ohno}},\
  }\href {\doibase 10.1103/RevModPhys.86.187} {\bibfield  {journal} {\bibinfo
  {journal} {Rev. Mod. Phys.}\ }\textbf {\bibinfo {volume} {86}},\ \bibinfo
  {pages} {187} (\bibinfo {year} {2014})}\BibitemShut {NoStop}%
\bibitem [{\citenamefont {Dietl}(2007)}]{Dietl00}%
  \BibitemOpen
  \bibfield  {author} {\bibinfo {author} {\bibfnamefont {T.}~\bibnamefont
  {Dietl}},\ }\href {http://stacks.iop.org/0953-8984/19/i=16/a=165204}
  {\bibfield  {journal} {\bibinfo  {journal} {Journal of Physics: Condensed
  Matter}\ }\textbf {\bibinfo {volume} {19}},\ \bibinfo {pages} {165204}
  (\bibinfo {year} {2007})}\BibitemShut {NoStop}%
\bibitem [{\citenamefont {Dietl}(2003)}]{Dietl:2003aa}%
  \BibitemOpen
  \bibfield  {author} {\bibinfo {author} {\bibfnamefont {T.}~\bibnamefont
  {Dietl}},\ }\href {http://dx.doi.org/10.1038/nmat989} {\bibfield  {journal}
  {\bibinfo  {journal} {Nat Mater}\ }\textbf {\bibinfo {volume} {2}},\ \bibinfo
  {pages} {646} (\bibinfo {year} {2003})}\BibitemShut {NoStop}%
\bibitem [{\citenamefont {Liu}\ and\ \citenamefont {Jacquier}(2007)}]{LiuYAG}%
  \BibitemOpen
  \bibinfo {editor} {\bibfnamefont {G.}~\bibnamefont {Liu}}\ and\ \bibinfo
  {editor} {\bibfnamefont {B.}~\bibnamefont {Jacquier}},\ eds.,\ \enquote
  {\bibinfo {title} {Spectroscopic properties of rare earths in optical
  materials},}\ \ (\bibinfo  {publisher} {Springer and Tsinghua University
  Press},\ \bibinfo {year} {2007})\ Chap.~\bibinfo {chapter} {7}\BibitemShut
  {NoStop}%
\bibitem [{\citenamefont {Poole}\ and\ \citenamefont {Farach}(1999)}]{PF99}%
  \BibitemOpen
  \bibinfo {editor} {\bibfnamefont {C.~P.~J.}\ \bibnamefont {Poole}}\ and\
  \bibinfo {editor} {\bibfnamefont {H.~A.}\ \bibnamefont {Farach}},\ eds.,\
  \href@noop {} {\emph {\bibinfo {title} {Handbook of Electron Spin
  Resonance}}},\ Vol.~\bibinfo {volume} {2}\ (\bibinfo  {publisher}
  {Springer-Verlag},\ \bibinfo {year} {1999})\BibitemShut {NoStop}%
\bibitem [{\citenamefont {Kaminskii}(1981)}]{raey}%
  \BibitemOpen
  \bibfield  {author} {\bibinfo {author} {\bibfnamefont {A.}~\bibnamefont
  {Kaminskii}},\ }in\ \href {\doibase 10.1007/978-3-540-34838-2_4} {\emph
  {\bibinfo {booktitle} {Laser Crystals}}},\ \bibinfo {series} {Springer Series
  in Optical Sciences}, Vol.~\bibinfo {volume} {14}\ (\bibinfo  {publisher}
  {Springer Berlin Heidelberg},\ \bibinfo {year} {1981})\ pp.\ \bibinfo {pages}
  {114--165}\BibitemShut {NoStop}%
\bibitem [{\citenamefont {O'Brien}\ \emph {et~al.}(2014)\citenamefont
  {O'Brien}, \citenamefont {Lauk}, \citenamefont {Blum}, \citenamefont
  {Morigi},\ and\ \citenamefont {Fleischhauer}}]{PhysRevLett.113.063603}%
  \BibitemOpen
  \bibfield  {author} {\bibinfo {author} {\bibfnamefont {C.}~\bibnamefont
  {O'Brien}}, \bibinfo {author} {\bibfnamefont {N.}~\bibnamefont {Lauk}},
  \bibinfo {author} {\bibfnamefont {S.}~\bibnamefont {Blum}}, \bibinfo {author}
  {\bibfnamefont {G.}~\bibnamefont {Morigi}}, \ and\ \bibinfo {author}
  {\bibfnamefont {M.}~\bibnamefont {Fleischhauer}},\ }\href {\doibase
  10.1103/PhysRevLett.113.063603} {\bibfield  {journal} {\bibinfo  {journal}
  {Phys. Rev. Lett.}\ }\textbf {\bibinfo {volume} {113}},\ \bibinfo {pages}
  {063603} (\bibinfo {year} {2014})}\BibitemShut {NoStop}%
\bibitem [{\citenamefont {Williamson}\ \emph {et~al.}(2014)\citenamefont
  {Williamson}, \citenamefont {Chen},\ and\ \citenamefont
  {Longdell}}]{Williamson:2014aa}%
  \BibitemOpen
  \bibfield  {author} {\bibinfo {author} {\bibfnamefont {L.~A.}\ \bibnamefont
  {Williamson}}, \bibinfo {author} {\bibfnamefont {Y.-H.}\ \bibnamefont
  {Chen}}, \ and\ \bibinfo {author} {\bibfnamefont {J.~J.}\ \bibnamefont
  {Longdell}},\ }\href {http://link.aps.org/doi/10.1103/PhysRevLett.113.203601}
  {\bibfield  {journal} {\bibinfo  {journal} {Physical Review Letters}\
  }\textbf {\bibinfo {volume} {113}},\ \bibinfo {pages} {203601} (\bibinfo
  {year} {2014})}\BibitemShut {NoStop}%
\bibitem [{\citenamefont {Tian}\ \emph {et~al.}(2004)\citenamefont {Tian},
  \citenamefont {Rabl}, \citenamefont {Blatt},\ and\ \citenamefont
  {Zoller}}]{PhysRevLett.92.247902}%
  \BibitemOpen
  \bibfield  {author} {\bibinfo {author} {\bibfnamefont {L.}~\bibnamefont
  {Tian}}, \bibinfo {author} {\bibfnamefont {P.}~\bibnamefont {Rabl}}, \bibinfo
  {author} {\bibfnamefont {R.}~\bibnamefont {Blatt}}, \ and\ \bibinfo {author}
  {\bibfnamefont {P.}~\bibnamefont {Zoller}},\ }\href {\doibase
  10.1103/PhysRevLett.92.247902} {\bibfield  {journal} {\bibinfo  {journal}
  {Phys. Rev. Lett.}\ }\textbf {\bibinfo {volume} {92}},\ \bibinfo {pages}
  {247902} (\bibinfo {year} {2004})}\BibitemShut {NoStop}%
\bibitem [{\citenamefont {Stannigel}\ \emph {et~al.}(2010)\citenamefont
  {Stannigel}, \citenamefont {Rabl}, \citenamefont {S\o{}rensen}, \citenamefont
  {Zoller},\ and\ \citenamefont {Lukin}}]{PhysRevLett.105.220501}%
  \BibitemOpen
  \bibfield  {author} {\bibinfo {author} {\bibfnamefont {K.}~\bibnamefont
  {Stannigel}}, \bibinfo {author} {\bibfnamefont {P.}~\bibnamefont {Rabl}},
  \bibinfo {author} {\bibfnamefont {A.~S.}\ \bibnamefont {S\o{}rensen}},
  \bibinfo {author} {\bibfnamefont {P.}~\bibnamefont {Zoller}}, \ and\ \bibinfo
  {author} {\bibfnamefont {M.~D.}\ \bibnamefont {Lukin}},\ }\href {\doibase
  10.1103/PhysRevLett.105.220501} {\bibfield  {journal} {\bibinfo  {journal}
  {Phys. Rev. Lett.}\ }\textbf {\bibinfo {volume} {105}},\ \bibinfo {pages}
  {220501} (\bibinfo {year} {2010})}\BibitemShut {NoStop}%
\bibitem [{\citenamefont {Stephens}\ \emph {et~al.}(2013)\citenamefont
  {Stephens}, \citenamefont {Huang}, \citenamefont {Nemoto},\ and\
  \citenamefont {Munro}}]{PhysRevA.87.052333}%
  \BibitemOpen
  \bibfield  {author} {\bibinfo {author} {\bibfnamefont {A.~M.}\ \bibnamefont
  {Stephens}}, \bibinfo {author} {\bibfnamefont {J.}~\bibnamefont {Huang}},
  \bibinfo {author} {\bibfnamefont {K.}~\bibnamefont {Nemoto}}, \ and\ \bibinfo
  {author} {\bibfnamefont {W.~J.}\ \bibnamefont {Munro}},\ }\href {\doibase
  10.1103/PhysRevA.87.052333} {\bibfield  {journal} {\bibinfo  {journal} {Phys.
  Rev. A}\ }\textbf {\bibinfo {volume} {87}},\ \bibinfo {pages} {052333}
  (\bibinfo {year} {2013})}\BibitemShut {NoStop}%
\bibitem [{\citenamefont {Strekalov}\ \emph {et~al.}(2009)\citenamefont
  {Strekalov}, \citenamefont {Schwefel}, \citenamefont {Savchenkov},
  \citenamefont {Matsko}, \citenamefont {Wang},\ and\ \citenamefont
  {Yu}}]{PhysRevA.80.033810}%
  \BibitemOpen
  \bibfield  {author} {\bibinfo {author} {\bibfnamefont {D.~V.}\ \bibnamefont
  {Strekalov}}, \bibinfo {author} {\bibfnamefont {H.~G.~L.}\ \bibnamefont
  {Schwefel}}, \bibinfo {author} {\bibfnamefont {A.~A.}\ \bibnamefont
  {Savchenkov}}, \bibinfo {author} {\bibfnamefont {A.~B.}\ \bibnamefont
  {Matsko}}, \bibinfo {author} {\bibfnamefont {L.~J.}\ \bibnamefont {Wang}}, \
  and\ \bibinfo {author} {\bibfnamefont {N.}~\bibnamefont {Yu}},\ }\href
  {\doibase 10.1103/PhysRevA.80.033810} {\bibfield  {journal} {\bibinfo
  {journal} {Phys. Rev. A}\ }\textbf {\bibinfo {volume} {80}},\ \bibinfo
  {pages} {033810} (\bibinfo {year} {2009})}\BibitemShut {NoStop}%
\bibitem [{\citenamefont {Goryachev}\ \emph {et~al.}(2013)\citenamefont
  {Goryachev}, \citenamefont {Farr},\ and\ \citenamefont
  {Tobar}}]{goryachev2013giant}%
  \BibitemOpen
  \bibfield  {author} {\bibinfo {author} {\bibfnamefont {M.}~\bibnamefont
  {Goryachev}}, \bibinfo {author} {\bibfnamefont {W.~G.}\ \bibnamefont {Farr}},
  \ and\ \bibinfo {author} {\bibfnamefont {M.~E.}\ \bibnamefont {Tobar}},\
  }\href@noop {} {\bibfield  {journal} {\bibinfo  {journal} {Applied Physics
  Letters}\ }\textbf {\bibinfo {volume} {103}},\ \bibinfo {pages} {262404}
  (\bibinfo {year} {2013})}\BibitemShut {NoStop}%
\bibitem [{\citenamefont {Farr}\ \emph {et~al.}(2014)\citenamefont {Farr},
  \citenamefont {Goryachev}, \citenamefont {Creedon},\ and\ \citenamefont
  {Tobar}}]{PhysRevB.90.054409}%
  \BibitemOpen
  \bibfield  {author} {\bibinfo {author} {\bibfnamefont {W.~G.}\ \bibnamefont
  {Farr}}, \bibinfo {author} {\bibfnamefont {M.}~\bibnamefont {Goryachev}},
  \bibinfo {author} {\bibfnamefont {D.~L.}\ \bibnamefont {Creedon}}, \ and\
  \bibinfo {author} {\bibfnamefont {M.~E.}\ \bibnamefont {Tobar}},\ }\href
  {\doibase 10.1103/PhysRevB.90.054409} {\bibfield  {journal} {\bibinfo
  {journal} {Phys. Rev. B}\ }\textbf {\bibinfo {volume} {90}},\ \bibinfo
  {pages} {054409} (\bibinfo {year} {2014})}\BibitemShut {NoStop}%
\bibitem [{\citenamefont {Le~Floch}\ \emph {et~al.}(2013)\citenamefont
  {Le~Floch}, \citenamefont {Fan}, \citenamefont {Aubourg}, \citenamefont
  {Cros}, \citenamefont {Carvalho}, \citenamefont {Shan}, \citenamefont
  {Bourhill}, \citenamefont {Ivanov}, \citenamefont {Humbert}, \citenamefont
  {Madrangeas},\ and\ \citenamefont {Tobar}}]{Le-Floch:2013aa}%
  \BibitemOpen
  \bibfield  {author} {\bibinfo {author} {\bibfnamefont {J.-M.}\ \bibnamefont
  {Le~Floch}}, \bibinfo {author} {\bibfnamefont {Y.}~\bibnamefont {Fan}},
  \bibinfo {author} {\bibfnamefont {M.}~\bibnamefont {Aubourg}}, \bibinfo
  {author} {\bibfnamefont {D.}~\bibnamefont {Cros}}, \bibinfo {author}
  {\bibfnamefont {N.~C.}\ \bibnamefont {Carvalho}}, \bibinfo {author}
  {\bibfnamefont {Q.}~\bibnamefont {Shan}}, \bibinfo {author} {\bibfnamefont
  {J.}~\bibnamefont {Bourhill}}, \bibinfo {author} {\bibfnamefont {E.~N.}\
  \bibnamefont {Ivanov}}, \bibinfo {author} {\bibfnamefont {G.}~\bibnamefont
  {Humbert}}, \bibinfo {author} {\bibfnamefont {V.}~\bibnamefont {Madrangeas}},
  \ and\ \bibinfo {author} {\bibfnamefont {M.~E.}\ \bibnamefont {Tobar}},\
  }\href
  {http://scitation.aip.org/content/aip/journal/rsi/84/12/10.1063/1.4848935}
  {\bibfield  {journal} {\bibinfo  {journal} {Review of Scientific
  Instruments}\ }\textbf {\bibinfo {volume} {84}},\  (\bibinfo {year}
  {2013})}\BibitemShut {NoStop}%
\bibitem [{sup()}]{suppm}%
  \BibitemOpen
  \href@noop {} {\enquote {\bibinfo {title} {See supplemental material at url
  for mode and coupling parameters and hysteresis characteristics},}\ }\bibinfo
  {howpublished} {URL}\BibitemShut {NoStop}%
\bibitem [{\citenamefont {Tobar}\ \emph {et~al.}(2001)\citenamefont {Tobar},
  \citenamefont {Hartnett}, \citenamefont {Ivanov}, \citenamefont {Blondy},\
  and\ \citenamefont {Cros}}]{mikehybridmode2001}%
  \BibitemOpen
  \bibfield  {author} {\bibinfo {author} {\bibfnamefont {M.}~\bibnamefont
  {Tobar}}, \bibinfo {author} {\bibfnamefont {J.}~\bibnamefont {Hartnett}},
  \bibinfo {author} {\bibfnamefont {E.}~\bibnamefont {Ivanov}}, \bibinfo
  {author} {\bibfnamefont {P.}~\bibnamefont {Blondy}}, \ and\ \bibinfo {author}
  {\bibfnamefont {D.}~\bibnamefont {Cros}},\ }\href {\doibase
  10.1109/19.918182} {\bibfield  {journal} {\bibinfo  {journal}
  {Instrumentation and Measurement, IEEE Transactions on}\ }\textbf {\bibinfo
  {volume} {50}},\ \bibinfo {pages} {522} (\bibinfo {year} {2001})}\BibitemShut
  {NoStop}%
\bibitem [{\citenamefont {Hartnett}\ \emph {et~al.}(2002)\citenamefont
  {Hartnett}, \citenamefont {Luiten}, \citenamefont {Krupka}, \citenamefont
  {Tobar},\ and\ \citenamefont {Bilski}}]{0022-3727-35-13-301}%
  \BibitemOpen
  \bibfield  {author} {\bibinfo {author} {\bibfnamefont {J.~G.}\ \bibnamefont
  {Hartnett}}, \bibinfo {author} {\bibfnamefont {A.~N.}\ \bibnamefont
  {Luiten}}, \bibinfo {author} {\bibfnamefont {J.}~\bibnamefont {Krupka}},
  \bibinfo {author} {\bibfnamefont {M.~E.}\ \bibnamefont {Tobar}}, \ and\
  \bibinfo {author} {\bibfnamefont {P.}~\bibnamefont {Bilski}},\ }\href
  {http://stacks.iop.org/0022-3727/35/i=13/a=301} {\bibfield  {journal}
  {\bibinfo  {journal} {Journal of Physics D: Applied Physics}\ }\textbf
  {\bibinfo {volume} {35}},\ \bibinfo {pages} {1459} (\bibinfo {year}
  {2002})}\BibitemShut {NoStop}%
\bibitem [{\citenamefont {Sun}\ \emph {et~al.}(2008)\citenamefont {Sun},
  \citenamefont {B\"ottger}, \citenamefont {Thiel},\ and\ \citenamefont
  {Cone}}]{PhysRevB.77.085124}%
  \BibitemOpen
  \bibfield  {author} {\bibinfo {author} {\bibfnamefont {Y.}~\bibnamefont
  {Sun}}, \bibinfo {author} {\bibfnamefont {T.}~\bibnamefont {B\"ottger}},
  \bibinfo {author} {\bibfnamefont {C.~W.}\ \bibnamefont {Thiel}}, \ and\
  \bibinfo {author} {\bibfnamefont {R.~L.}\ \bibnamefont {Cone}},\ }\href
  {\doibase 10.1103/PhysRevB.77.085124} {\bibfield  {journal} {\bibinfo
  {journal} {Phys. Rev. B}\ }\textbf {\bibinfo {volume} {77}},\ \bibinfo
  {pages} {085124} (\bibinfo {year} {2008})}\BibitemShut {NoStop}%
\bibitem [{\citenamefont {Probst}\ \emph {et~al.}(2014)\citenamefont {Probst},
  \citenamefont {Tkal\ifmmode~\check{c}\else \v{c}\fi{}ec}, \citenamefont
  {Rotzinger}, \citenamefont {Rieger}, \citenamefont {Le~Floch}, \citenamefont
  {Goryachev}, \citenamefont {Tobar}, \citenamefont {Ustinov},\ and\
  \citenamefont {Bushev}}]{PhysRevB.90.100404}%
  \BibitemOpen
  \bibfield  {author} {\bibinfo {author} {\bibfnamefont {S.}~\bibnamefont
  {Probst}}, \bibinfo {author} {\bibfnamefont {A.}~\bibnamefont
  {Tkal\ifmmode~\check{c}\else \v{c}\fi{}ec}}, \bibinfo {author} {\bibfnamefont
  {H.}~\bibnamefont {Rotzinger}}, \bibinfo {author} {\bibfnamefont
  {D.}~\bibnamefont {Rieger}}, \bibinfo {author} {\bibfnamefont {J.-M.}\
  \bibnamefont {Le~Floch}}, \bibinfo {author} {\bibfnamefont {M.}~\bibnamefont
  {Goryachev}}, \bibinfo {author} {\bibfnamefont {M.~E.}\ \bibnamefont
  {Tobar}}, \bibinfo {author} {\bibfnamefont {A.~V.}\ \bibnamefont {Ustinov}},
  \ and\ \bibinfo {author} {\bibfnamefont {P.~A.}\ \bibnamefont {Bushev}},\
  }\href {\doibase 10.1103/PhysRevB.90.100404} {\bibfield  {journal} {\bibinfo
  {journal} {Phys. Rev. B}\ }\textbf {\bibinfo {volume} {90}},\ \bibinfo
  {pages} {100404} (\bibinfo {year} {2014})}\BibitemShut {NoStop}%
\bibitem [{\citenamefont {Bushev}\ \emph {et~al.}(2011)\citenamefont {Bushev},
  \citenamefont {Feofanov}, \citenamefont {Rotzinger}, \citenamefont
  {Protopopov}, \citenamefont {Cole}, \citenamefont {Wilson}, \citenamefont
  {Fischer}, \citenamefont {Lukashenko},\ and\ \citenamefont
  {Ustinov}}]{PhysRevB.84.060501}%
  \BibitemOpen
  \bibfield  {author} {\bibinfo {author} {\bibfnamefont {P.}~\bibnamefont
  {Bushev}}, \bibinfo {author} {\bibfnamefont {A.~K.}\ \bibnamefont
  {Feofanov}}, \bibinfo {author} {\bibfnamefont {H.}~\bibnamefont {Rotzinger}},
  \bibinfo {author} {\bibfnamefont {I.}~\bibnamefont {Protopopov}}, \bibinfo
  {author} {\bibfnamefont {J.~H.}\ \bibnamefont {Cole}}, \bibinfo {author}
  {\bibfnamefont {C.~M.}\ \bibnamefont {Wilson}}, \bibinfo {author}
  {\bibfnamefont {G.}~\bibnamefont {Fischer}}, \bibinfo {author} {\bibfnamefont
  {A.}~\bibnamefont {Lukashenko}}, \ and\ \bibinfo {author} {\bibfnamefont
  {A.~V.}\ \bibnamefont {Ustinov}},\ }\href {\doibase
  10.1103/PhysRevB.84.060501} {\bibfield  {journal} {\bibinfo  {journal} {Phys.
  Rev. B}\ }\textbf {\bibinfo {volume} {84}},\ \bibinfo {pages} {060501}
  (\bibinfo {year} {2011})}\BibitemShut {NoStop}%
\bibitem [{\citenamefont {Sandner}\ \emph {et~al.}(2012)\citenamefont
  {Sandner}, \citenamefont {Ritsch}, \citenamefont {Ams{\"u}ss}, \citenamefont
  {Koller}, \citenamefont {N{\"o}bauer}, \citenamefont {Putz}, \citenamefont
  {Schmiedmayer},\ and\ \citenamefont {Majer}}]{Sandner:2012aa}%
  \BibitemOpen
  \bibfield  {author} {\bibinfo {author} {\bibfnamefont {K.}~\bibnamefont
  {Sandner}}, \bibinfo {author} {\bibfnamefont {H.}~\bibnamefont {Ritsch}},
  \bibinfo {author} {\bibfnamefont {R.}~\bibnamefont {Ams{\"u}ss}}, \bibinfo
  {author} {\bibfnamefont {C.}~\bibnamefont {Koller}}, \bibinfo {author}
  {\bibfnamefont {T.}~\bibnamefont {N{\"o}bauer}}, \bibinfo {author}
  {\bibfnamefont {S.}~\bibnamefont {Putz}}, \bibinfo {author} {\bibfnamefont
  {J.}~\bibnamefont {Schmiedmayer}}, \ and\ \bibinfo {author} {\bibfnamefont
  {J.}~\bibnamefont {Majer}},\ }\href
  {http://link.aps.org/doi/10.1103/PhysRevA.85.053806} {\bibfield  {journal}
  {\bibinfo  {journal} {Physical Review A}\ }\textbf {\bibinfo {volume} {85}},\
  \bibinfo {pages} {053806} (\bibinfo {year} {2012})}\BibitemShut {NoStop}%
\bibitem [{\citenamefont {Goryachev}\ \emph
  {et~al.}(2014{\natexlab{c}})\citenamefont {Goryachev}, \citenamefont {Farr},
  \citenamefont {Creedon},\ and\ \citenamefont {Tobar}}]{PhysRevB.89.224407}%
  \BibitemOpen
  \bibfield  {author} {\bibinfo {author} {\bibfnamefont {M.}~\bibnamefont
  {Goryachev}}, \bibinfo {author} {\bibfnamefont {W.~G.}\ \bibnamefont {Farr}},
  \bibinfo {author} {\bibfnamefont {D.~L.}\ \bibnamefont {Creedon}}, \ and\
  \bibinfo {author} {\bibfnamefont {M.~E.}\ \bibnamefont {Tobar}},\ }\href
  {\doibase 10.1103/PhysRevB.89.224407} {\bibfield  {journal} {\bibinfo
  {journal} {Phys. Rev. B}\ }\textbf {\bibinfo {volume} {89}},\ \bibinfo
  {pages} {224407} (\bibinfo {year} {2014}{\natexlab{c}})}\BibitemShut
  {NoStop}%
\bibitem [{\citenamefont {Krupka}\ \emph {et~al.}(1999)\citenamefont {Krupka},
  \citenamefont {Derzakowski}, \citenamefont {Tobar}, \citenamefont
  {Hartnett},\ and\ \citenamefont {Geyer}}]{krupka1999}%
  \BibitemOpen
  \bibfield  {author} {\bibinfo {author} {\bibfnamefont {J.}~\bibnamefont
  {Krupka}}, \bibinfo {author} {\bibfnamefont {K.}~\bibnamefont {Derzakowski}},
  \bibinfo {author} {\bibfnamefont {M.}~\bibnamefont {Tobar}}, \bibinfo
  {author} {\bibfnamefont {J.}~\bibnamefont {Hartnett}}, \ and\ \bibinfo
  {author} {\bibfnamefont {R.~G.}\ \bibnamefont {Geyer}},\ }\href
  {http://stacks.iop.org/0957-0233/10/i=5/a=308} {\bibfield  {journal}
  {\bibinfo  {journal} {Measurement Science and Technology}\ }\textbf {\bibinfo
  {volume} {10}},\ \bibinfo {pages} {387} (\bibinfo {year} {1999})}\BibitemShut
  {NoStop}%
\bibitem [{\citenamefont {Hartnett}\ \emph {et~al.}(2011)\citenamefont
  {Hartnett}, \citenamefont {Jaeckel}, \citenamefont {Povey},\ and\
  \citenamefont {Tobar}}]{Hartnett:2011aa}%
  \BibitemOpen
  \bibfield  {author} {\bibinfo {author} {\bibfnamefont {J.~G.}\ \bibnamefont
  {Hartnett}}, \bibinfo {author} {\bibfnamefont {J.}~\bibnamefont {Jaeckel}},
  \bibinfo {author} {\bibfnamefont {R.~G.}\ \bibnamefont {Povey}}, \ and\
  \bibinfo {author} {\bibfnamefont {M.~E.}\ \bibnamefont {Tobar}},\ }\href
  {\doibase http://dx.doi.org/10.1016/j.physletb.2011.03.022} {\bibfield
  {journal} {\bibinfo  {journal} {Physics Letters B}\ }\textbf {\bibinfo
  {volume} {698}},\ \bibinfo {pages} {346} (\bibinfo {year}
  {2011})}\BibitemShut {NoStop}%
\bibitem [{\citenamefont {Tavis}\ and\ \citenamefont
  {Cummings}(1968)}]{PhysRev.170.379}%
  \BibitemOpen
  \bibfield  {author} {\bibinfo {author} {\bibfnamefont {M.}~\bibnamefont
  {Tavis}}\ and\ \bibinfo {author} {\bibfnamefont {F.~W.}\ \bibnamefont
  {Cummings}},\ }\href {\doibase 10.1103/PhysRev.170.379} {\bibfield  {journal}
  {\bibinfo  {journal} {Phys. Rev.}\ }\textbf {\bibinfo {volume} {170}},\
  \bibinfo {pages} {379} (\bibinfo {year} {1968})}\BibitemShut {NoStop}%
\bibitem [{\citenamefont {TAVIS}\ and\ \citenamefont
  {CUMMINGS}(1969)}]{PhysRev.188.692}%
  \BibitemOpen
  \bibfield  {author} {\bibinfo {author} {\bibfnamefont {M.}~\bibnamefont
  {TAVIS}}\ and\ \bibinfo {author} {\bibfnamefont {F.~W.}\ \bibnamefont
  {CUMMINGS}},\ }\href {\doibase 10.1103/PhysRev.188.692} {\bibfield  {journal}
  {\bibinfo  {journal} {Phys. Rev.}\ }\textbf {\bibinfo {volume} {188}},\
  \bibinfo {pages} {692} (\bibinfo {year} {1969})}\BibitemShut {NoStop}%
\bibitem [{\citenamefont {Staudt}\ \emph {et~al.}(2012)\citenamefont {Staudt},
  \citenamefont {Hoi}, \citenamefont {Krantz}, \citenamefont {Sandberg},
  \citenamefont {Simoen}, \citenamefont {Bushev}, \citenamefont {Sangouard},
  \citenamefont {Afzelius}, \citenamefont {Shumeiko}, \citenamefont
  {Johansson}, \citenamefont {Delsing},\ and\ \citenamefont
  {Wilson}}]{Staudt:2012aa}%
  \BibitemOpen
  \bibfield  {author} {\bibinfo {author} {\bibfnamefont {M.~U.}\ \bibnamefont
  {Staudt}}, \bibinfo {author} {\bibfnamefont {I.-C.}\ \bibnamefont {Hoi}},
  \bibinfo {author} {\bibfnamefont {P.}~\bibnamefont {Krantz}}, \bibinfo
  {author} {\bibfnamefont {M.}~\bibnamefont {Sandberg}}, \bibinfo {author}
  {\bibfnamefont {M.}~\bibnamefont {Simoen}}, \bibinfo {author} {\bibfnamefont
  {P.}~\bibnamefont {Bushev}}, \bibinfo {author} {\bibfnamefont
  {N.}~\bibnamefont {Sangouard}}, \bibinfo {author} {\bibfnamefont
  {M.}~\bibnamefont {Afzelius}}, \bibinfo {author} {\bibfnamefont {V.~S.}\
  \bibnamefont {Shumeiko}}, \bibinfo {author} {\bibfnamefont {G.}~\bibnamefont
  {Johansson}}, \bibinfo {author} {\bibfnamefont {P.}~\bibnamefont {Delsing}},
  \ and\ \bibinfo {author} {\bibfnamefont {C.~M.}\ \bibnamefont {Wilson}},\
  }\href {http://stacks.iop.org/0953-4075/45/i=12/a=124019} {\bibfield
  {journal} {\bibinfo  {journal} {Journal of Physics B: Atomic, Molecular and
  Optical Physics}\ }\textbf {\bibinfo {volume} {45}},\ \bibinfo {pages}
  {124019} (\bibinfo {year} {2012})}\BibitemShut {NoStop}%
\bibitem [{\citenamefont {Baibekov}\ \emph {et~al.}(2011)\citenamefont
  {Baibekov}, \citenamefont {Kurkin}, \citenamefont {Gafurov}, \citenamefont
  {Endeward}, \citenamefont {Rakhmatullin},\ and\ \citenamefont
  {Mamin}}]{Baibekov:2011aa}%
  \BibitemOpen
  \bibfield  {author} {\bibinfo {author} {\bibfnamefont {E.}~\bibnamefont
  {Baibekov}}, \bibinfo {author} {\bibfnamefont {I.}~\bibnamefont {Kurkin}},
  \bibinfo {author} {\bibfnamefont {M.}~\bibnamefont {Gafurov}}, \bibinfo
  {author} {\bibfnamefont {B.}~\bibnamefont {Endeward}}, \bibinfo {author}
  {\bibfnamefont {R.}~\bibnamefont {Rakhmatullin}}, \ and\ \bibinfo {author}
  {\bibfnamefont {G.}~\bibnamefont {Mamin}},\ }\href {\doibase
  http://dx.doi.org/10.1016/j.jmr.2010.12.015} {\bibfield  {journal} {\bibinfo
  {journal} {Journal of Magnetic Resonance}\ }\textbf {\bibinfo {volume}
  {209}},\ \bibinfo {pages} {61} (\bibinfo {year} {2011})}\BibitemShut
  {NoStop}%
\bibitem [{\citenamefont {Guillot-No\"el}\ \emph {et~al.}(2000)\citenamefont
  {Guillot-No\"el}, \citenamefont {Mehta}, \citenamefont {Viana}, \citenamefont
  {Gourier}, \citenamefont {Boukhris},\ and\ \citenamefont
  {Jandl}}]{PhysRevB.61.15338}%
  \BibitemOpen
  \bibfield  {author} {\bibinfo {author} {\bibfnamefont {O.}~\bibnamefont
  {Guillot-No\"el}}, \bibinfo {author} {\bibfnamefont {V.}~\bibnamefont
  {Mehta}}, \bibinfo {author} {\bibfnamefont {B.}~\bibnamefont {Viana}},
  \bibinfo {author} {\bibfnamefont {D.}~\bibnamefont {Gourier}}, \bibinfo
  {author} {\bibfnamefont {M.}~\bibnamefont {Boukhris}}, \ and\ \bibinfo
  {author} {\bibfnamefont {S.}~\bibnamefont {Jandl}},\ }\href {\doibase
  10.1103/PhysRevB.61.15338} {\bibfield  {journal} {\bibinfo  {journal} {Phys.
  Rev. B}\ }\textbf {\bibinfo {volume} {61}},\ \bibinfo {pages} {15338}
  (\bibinfo {year} {2000})}\BibitemShut {NoStop}%
\end{thebibliography}

%

\end{document}